\documentclass[11pt,letterpaper]{article}
\usepackage{fullpage,times}
\ifx\undefined\psfig\else \fi

\edef\psfigRestoreAt{\catcode`@=\number\catcode`@\relax}
\catcode`\@=11\relax
\newwrite\@unused
\def\typeout#1{{\let\protect\string\immediate\write\@unused{#1}}}
\def\figurepath{./}

\def\@nnil{\@nil}
\def\@empty{}
\def\@psdonoop#1\@@#2#3{}
\def\@psdo#1:=#2\do#3{\edef\@psdotmp{#2}\ifx\@psdotmp\@empty \else
    \expandafter\@psdoloop#2,\@nil,\@nil\@@#1{#3}\fi}
\def\@psdoloop#1,#2,#3\@@#4#5{\def#4{#1}\ifx #4\@nnil \else
       #5\def#4{#2}\ifx #4\@nnil \else#5\@ipsdoloop #3\@@#4{#5}\fi\fi}
\def\@ipsdoloop#1,#2\@@#3#4{\def#3{#1}\ifx #3\@nnil 
       \let\@nextwhile=\@psdonoop \else
      #4\relax\let\@nextwhile=\@ipsdoloop\fi\@nextwhile#2\@@#3{#4}}
\def\@tpsdo#1:=#2\do#3{\xdef\@psdotmp{#2}\ifx\@psdotmp\@empty \else
    \@tpsdoloop#2\@nil\@nil\@@#1{#3}\fi}
\def\@tpsdoloop#1#2\@@#3#4{\def#3{#1}\ifx #3\@nnil 
       \let\@nextwhile=\@psdonoop \else
      #4\relax\let\@nextwhile=\@tpsdoloop\fi\@nextwhile#2\@@#3{#4}}
\newread\ps@stream
\newif\ifnot@eof       
\newif\if@noisy        
\newif\if@atend        
\newif\if@psfile       
{\catcode`\%=12\global\gdef\epsf@start{
\def\epsf@PS{PS}
\def\epsf@getbb#1{%
\openin\ps@stream=#1
\ifeof\ps@stream\typeout{Error, File #1 not found}\else
   {\not@eoftrue \chardef\other=12
    \def\do##1{\catcode`##1=\other}\dospecials \catcode`\ =10
    \loop
       \if@psfile
	  \read\ps@stream to \epsf@fileline
       \else{
	  \obeyspaces
          \read\ps@stream to \epsf@tmp\global\let\epsf@fileline\epsf@tmp}
       \fi
       \ifeof\ps@stream\not@eoffalse\else
       \if@psfile\else
       \expandafter\epsf@test\epsf@fileline:. \\%
       \fi
          \expandafter\epsf@aux\epsf@fileline:. \\%
       \fi
   \ifnot@eof\repeat
   }\closein\ps@stream\fi}%
\long\def\epsf@test#1#2#3:#4\\{\def\epsf@testit{#1#2}
			\ifx\epsf@testit\epsf@start\else
\typeout{Warning! File does not start with `\epsf@start'.  It may not be a PostScript file.}
			\fi
			\@psfiletrue} 
{\catcode`\%=12\global\let\epsf@percent=
\long\def\epsf@aux#1#2:#3\\{\ifx#1\epsf@percent
   \def\epsf@testit{#2}\ifx\epsf@testit\epsf@bblit
	\@atendfalse
        \epsf@atend #3 . \\%
	\if@atend	
	   \if@verbose{
		\typeout{psfig: found `(atend)'; continuing search}
	   }\fi
        \else
        \epsf@grab #3 . . . \\%
        \not@eoffalse
        \global\no@bbfalse
        \fi
   \fi\fi}%
\def\epsf@grab #1 #2 #3 #4 #5\\{%
   \global\def\epsf@llx{#1}\ifx\epsf@llx\empty
      \epsf@grab #2 #3 #4 #5 .\\\else
   \global\def\epsf@lly{#2}%
   \global\def\epsf@urx{#3}\global\def\epsf@ury{#4}\fi}%
\def\epsf@atendlit{(atend)} 
\def\epsf@atend #1 #2 #3\\{%
   \def\epsf@tmp{#1}\ifx\epsf@tmp\empty
      \epsf@atend #2 #3 .\\\else
   \ifx\epsf@tmp\epsf@atendlit\@atendtrue\fi\fi}

\chardef\letter = 11
\chardef\other = 12

\newif \ifdebug 
\newif\ifc@mpute 
\c@mputetrue 

\let\then = \relax
\def\r@dian{pt }
\let\r@dians = \r@dian
\let\dimensionless@nit = \r@dian
\let\dimensionless@nits = \dimensionless@nit
\def\internal@nit{sp }
\let\internal@nits = \internal@nit
\newif\ifstillc@nverging
\def \Mess@ge #1{\ifdebug \then \message {#1} \fi}

{ 
	\catcode `\@ = \letter
	\gdef \nodimen {\expandafter \n@dimen \the \dimen}
	\gdef \term #1 #2 #3%
	       {\edef \t@ {\the #1}
		\edef \t@@ {\expandafter \n@dimen \the #2\r@dian}%
		\t@rm {\t@} {\t@@} {#3}%
	       }
	\gdef \t@rm #1 #2 #3%
	       {{%
		\count 0 = 0
		\dimen 0 = 1 \dimensionless@nit
		\dimen 2 = #2\relax
		\Mess@ge {Calculating term #1 of \nodimen 2}%
		\loop
		\ifnum	\count 0 < #1
		\then	\advance \count 0 by 1
			\Mess@ge {Iteration \the \count 0 \space}%
			\Multiply \dimen 0 by {\dimen 2}%
			\Mess@ge {After multiplication, term = \nodimen 0}%
			\Divide \dimen 0 by {\count 0}%
			\Mess@ge {After division, term = \nodimen 0}%
		\repeat
		\Mess@ge {Final value for term #1 of 
				\nodimen 2 \space is \nodimen 0}%
		\xdef \Term {#3 = \nodimen 0 \r@dians}%
		\aftergroup \Term
	       }}
	\catcode `\p = \other
	\catcode `\t = \other
	\gdef \n@dimen #1pt{#1} 
}

\def \Divide #1by #2{\divide #1 by #2} 

\def \Multiply #1by #2
       {{
	\count 0 = #1\relax
	\count 2 = #2\relax
	\count 4 = 65536
	\Mess@ge {Before scaling, count 0 = \the \count 0 \space and
			count 2 = \the \count 2}%
	\ifnum	\count 0 > 32767 
	\then	\divide \count 0 by 4
		\divide \count 4 by 4
	\else	\ifnum	\count 0 < -32767
		\then	\divide \count 0 by 4
			\divide \count 4 by 4
		\else
		\fi
	\fi
	\ifnum	\count 2 > 32767 
	\then	\divide \count 2 by 4
		\divide \count 4 by 4
	\else	\ifnum	\count 2 < -32767
		\then	\divide \count 2 by 4
			\divide \count 4 by 4
		\else
		\fi
	\fi
	\multiply \count 0 by \count 2
	\divide \count 0 by \count 4
	\xdef \product {#1 = \the \count 0 \internal@nits}%
	\aftergroup \product
       }}

\def\r@duce{\ifdim\dimen0 > 90\r@dian \then   
		\multiply\dimen0 by -1
		\advance\dimen0 by 180\r@dian
		\r@duce
	    \else \ifdim\dimen0 < -90\r@dian \then  
		\advance\dimen0 by 360\r@dian
		\r@duce
		\fi
	    \fi}

\def\Sine#1%
       {{%
	\dimen 0 = #1 \r@dian
	\r@duce
	\ifdim\dimen0 = -90\r@dian \then
	   \dimen4 = -1\r@dian
	   \c@mputefalse
	\fi
	\ifdim\dimen0 = 90\r@dian \then
	   \dimen4 = 1\r@dian
	   \c@mputefalse
	\fi
	\ifdim\dimen0 = 0\r@dian \then
	   \dimen4 = 0\r@dian
	   \c@mputefalse
	\fi
	\ifc@mpute \then
		\divide\dimen0 by 180
		\dimen0=3.141592654\dimen0
		\dimen 2 = 3.1415926535897963\r@dian 
		\divide\dimen 2 by 2 
		\Mess@ge {Sin: calculating Sin of \nodimen 0}%
		\count 0 = 1 
		\dimen 2 = 1 \r@dian 
		\dimen 4 = 0 \r@dian 
		\loop
			\ifnum	\dimen 2 = 0 
			\then	\stillc@nvergingfalse 
			\else	\stillc@nvergingtrue
			\fi
			\ifstillc@nverging 
			\then	\term {\count 0} {\dimen 0} {\dimen 2}%
				\advance \count 0 by 2
				\count 2 = \count 0
				\divide \count 2 by 2
				\ifodd	\count 2 
				\then	\advance \dimen 4 by \dimen 2
				\else	\advance \dimen 4 by -\dimen 2
				\fi
		\repeat
	\fi		
			\xdef \sine {\nodimen 4}%
       }}

\def\Cosine#1{\ifx\sine\UnDefined\edef\Savesine{\relax}\else
		             \edef\Savesine{\sine}\fi
	{\dimen0=#1\r@dian\multiply\dimen0 by -1
	 \advance\dimen0 by 90\r@dian
	 \Sine{\nodimen 0}
	 \xdef\cosine{\sine}
	 \xdef\sine{\Savesine}}}	      

\def\psdraft{
	\def\@psdraft{0}
}
\def\psfull{
	\def\@psdraft{100}
}

\psfull

\newif\if@draftbox
\def\psnodraftbox{
	\@draftboxfalse
}
\@draftboxtrue

\newif\if@prologfile
\newif\if@postlogfile
\def\pssilent{
	\@noisyfalse
}
\def\psnoisy{
	\@noisytrue
}
\psnoisy
\newif\if@bbllx
\newif\if@bblly
\newif\if@bburx
\newif\if@bbury
\newif\if@height
\newif\if@width
\newif\if@rheight
\newif\if@rwidth
\newif\if@angle
\newif\if@clip
\newif\if@verbose
\newif\if@scale
\def\@p@@sclip#1{\@cliptrue}

\def\@p@@sfile#1{\def\@p@sfile{null}%
	        \openin1=#1
		\ifeof1\closein1%
		       \openin1=\figurepath#1
			\ifeof1\typeout{Error, File #1 not found}
			   \if@bbllx\if@bblly\if@bburx\if@bbury
			      \def\@p@sfile{#1}%
			   \fi\fi\fi\fi
			\else\closein1
			    \edef\@p@sfile{\figurepath#1}%
                        \fi%
		 \else\closein1%
		       \def\@p@sfile{#1}%
		 \fi}
\def\@p@@sfigure#1{\def\@p@sfile{null}%
	        \openin1=#1
		\ifeof1\closein1%
		       \openin1=\figurepath#1
			\ifeof1\typeout{Error, File #1 not found}
			   \if@bbllx\if@bblly\if@bburx\if@bbury
			      \def\@p@sfile{#1}%
			   \fi\fi\fi\fi
			\else\closein1
			    \def\@p@sfile{\figurepath#1}%
                        \fi%
		 \else\closein1%
		       \def\@p@sfile{#1}%
		 \fi}

\def\@p@@sbbllx#1{
		\@bbllxtrue
		\dimen100=#1
		\edef\@p@sbbllx{\number\dimen100}
}
\def\@p@@sbblly#1{
		\@bbllytrue
		\dimen100=#1
		\edef\@p@sbblly{\number\dimen100}
}
\def\@p@@sbburx#1{
		\@bburxtrue
		\dimen100=#1
		\edef\@p@sbburx{\number\dimen100}
}
\def\@p@@sbbury#1{
		\@bburytrue
		\dimen100=#1
		\edef\@p@sbbury{\number\dimen100}
}
\def\@p@@sheight#1{
		\@heighttrue
		\dimen100=#1
   		\edef\@p@sheight{\number\dimen100}
}
\def\@p@@swidth#1{
		\@widthtrue
		\dimen100=#1
		\edef\@p@swidth{\number\dimen100}
}
\def\@p@@srheight#1{
		\@rheighttrue
		\dimen100=#1
		\edef\@p@srheight{\number\dimen100}
}
\def\@p@@srwidth#1{
		\@rwidthtrue
		\dimen100=#1
		\edef\@p@srwidth{\number\dimen100}
}
\def\@p@@sangle#1{
		\@angletrue
		\edef\@p@sangle{#1} 
}
\def\@p@@ssilent#1{ 
		\@verbosefalse
}
\def\@p@@sscale#1{
		\def\@p@scale{#1}
		\@scaletrue
}
\def\@p@@sprolog#1{\@prologfiletrue\def\@prologfileval{#1}}
\def\@p@@spostlog#1{\@postlogfiletrue\def\@postlogfileval{#1}}
\def\@cs@name#1{\csname #1\endcsname}
\def\@setparms#1=#2,{\@cs@name{@p@@s#1}{#2}}
\def\ps@init@parms{
		\@bbllxfalse \@bbllyfalse
		\@bburxfalse \@bburyfalse
		\@heightfalse \@widthfalse
		\@rheightfalse \@rwidthfalse
		\@scalefalse
		\def\@p@sbbllx{}\def\@p@sbblly{}
		\def\@p@sbburx{}\def\@p@sbbury{}
		\def\@p@sheight{}\def\@p@swidth{}
		\def\@p@srheight{}\def\@p@srwidth{}
		\def\@p@sangle{0}
		\def\@p@sfile{}
		\def\@p@scost{10}
		\def\@sc{}
		\@prologfilefalse
		\@postlogfilefalse
		\@clipfalse
		\if@noisy
			\@verbosetrue
		\else
			\@verbosefalse
		\fi
}
\def\parse@ps@parms#1{
	 	\@psdo\@psfiga:=#1\do
		   {\expandafter\@setparms\@psfiga,}}
\newif\ifno@bb
\def\bb@missing{
	\if@verbose{
		\typeout{psfig: searching \@p@sfile \space  for bounding box}
	}\fi
	\no@bbtrue
	\epsf@getbb{\@p@sfile}
        \ifno@bb \else \bb@cull\epsf@llx\epsf@lly\epsf@urx\epsf@ury\fi
}	
\def\bb@cull#1#2#3#4{
	\dimen100=#1 bp\edef\@p@sbbllx{\number\dimen100}
	\dimen100=#2 bp\edef\@p@sbblly{\number\dimen100}
	\dimen100=#3 bp\edef\@p@sbburx{\number\dimen100}
	\dimen100=#4 bp\edef\@p@sbbury{\number\dimen100}
	\no@bbfalse
}

\newdimen\p@intvaluex
\newdimen\p@intvaluey
\newdimen\@ffsetvalue
\newdimen\x@ffsetvalue
\newdimen\y@ffsetvalue

\def\compute@offset#1#2{{\dimen0=#1 sp\dimen1=#2 sp
			\advance\dimen1 by -\dimen0
			\dimen1=\sine\dimen1
			\dimen0=\cosine\dimen1
			\ifdim\dimen0<0sp \dimen1=0sp \fi
			\global\@ffsetvalue=\dimen1}}

\def\rotate@#1#2{{\dimen0=#1 sp\dimen1=#2 sp
		  \global\p@intvaluex=\cosine\dimen0
		  \dimen3=\sine\dimen1
		  \global\advance\p@intvaluex by -\dimen3
		  \global\p@intvaluey=\sine\dimen0
		  \dimen3=\cosine\dimen1
		  \global\advance\p@intvaluey by \dimen3
		  }}
\def\compute@bb{
		\no@bbfalse
		\if@bbllx \else \no@bbtrue \fi
		\if@bblly \else \no@bbtrue \fi
		\if@bburx \else \no@bbtrue \fi
		\if@bbury \else \no@bbtrue \fi
		\ifno@bb \bb@missing \fi
		\ifno@bb \typeout{FATAL ERROR: no bb supplied or found}
			\no-bb-error
		\fi
		\if@angle 
			\Sine{\@p@sangle}\Cosine{\@p@sangle}
			\compute@offset{\@p@sbblly}{\@p@sbbury}
			\x@ffsetvalue=\@ffsetvalue
			\compute@offset{\@p@sbburx}{\@p@sbbllx}
			\y@ffsetvalue=\@ffsetvalue

			\rotate@{\@p@sbbllx}{\@p@sbblly}
			\advance\p@intvaluex by -\x@ffsetvalue
			\advance\p@intvaluey by -\y@ffsetvalue
			\edef\@p@sbbllx{\number\p@intvaluex}
			\edef\@p@sbblly{\number\p@intvaluey}

			\rotate@{\@p@sbburx}{\@p@sbbury}
			\advance\p@intvaluex by \x@ffsetvalue
			\advance\p@intvaluey by \y@ffsetvalue
			\edef\@p@sbburx{\number\p@intvaluex}
			\edef\@p@sbbury{\number\p@intvaluey}
			{
			 \count0=\@p@sbbllx \count1=\@p@sbblly
		 	 \count2=\@p@sbburx \count3=\@p@sbbury
			 \dimen0=\@p@sbbllx sp\dimen1=\@p@sbblly sp
		 	 \dimen2=\@p@sbburx sp\dimen3=\@p@sbbury sp
			 \dimen203=\dimen2 \advance\dimen203 by -\dimen0
			 \dimen204=\dimen3 \advance\dimen204 by -\dimen1
			 \ifdim\dimen203<0sp 
			      \count203=\count2 \count2=\count0 
			      \count0=\count203 
			      \global\edef\@p@sbbllx{\number\count0}
			      \global\edef\@p@sbburx{\number\count2}
			 \fi
			 \ifdim\dimen204<0sp 
			       \count204=\count3
			       \count3=\count1
			       \count1=\count204
			       \global\edef\@p@sbblly{\number\count1}
			       \global\edef\@p@sbbury{\number\count3}
			 \fi
			}
		\fi
		\count203=\@p@sbburx
		\count204=\@p@sbbury
		\advance\count203 by -\@p@sbbllx
		\advance\count204 by -\@p@sbblly
		\edef\@bbw{\number\count203}
		\edef\@bbh{\number\count204}
}
\def\in@hundreds#1#2#3{\count240=#2 \count241=#3
		     \count100=\count240	
		     \divide\count100 by \count241
		     \count101=\count100
		     \multiply\count101 by \count241
		     \advance\count240 by -\count101
		     \multiply\count240 by 10
		     \count101=\count240	
		     \divide\count101 by \count241
		     \count102=\count101
		     \multiply\count102 by \count241
		     \advance\count240 by -\count102
		     \multiply\count240 by 10
		     \count102=\count240	
		     \divide\count102 by \count241
		     \count200=#1\count205=0
		     \count201=\count200
			\multiply\count201 by \count100
		 	\advance\count205 by \count201
		     \count201=\count200
			\divide\count201 by 10
			\multiply\count201 by \count101
			\advance\count205 by \count201
		     \count201=\count200
			\divide\count201 by 100
			\multiply\count201 by \count102
			\advance\count205 by \count201
		     \edef\@result{\number\count205}
}
\def\@ScaleInHundreds#1{
		\in@hundreds{#1}{\@p@scale}{100}
		\edef#1{\@result}
}
\def\compute@wfromh{
		\in@hundreds{\@p@sheight}{\@bbw}{\@bbh}
		\edef\@p@swidth{\@result}
}
\def\compute@hfromw{
		\in@hundreds{\@p@swidth}{\@bbh}{\@bbw}
		\edef\@p@sheight{\@result}
}
\def\compute@handw{
		\if@height 
			\if@width
			\else
				\compute@wfromh
			\fi
		\else 
			\if@width
				\compute@hfromw
			\else
				\edef\@p@sheight{\@bbh}
				\edef\@p@swidth{\@bbw}
			\fi
		\fi
}
\def\compute@resv{
		\if@rheight \else \edef\@p@srheight{\@p@sheight} \fi
		\if@rwidth \else \edef\@p@srwidth{\@p@swidth} \fi
}
\def\compute@sizes{
	\compute@bb
	\compute@handw
	\compute@resv
}
\def\psfig#1{\vbox {
	\ps@init@parms
	\parse@ps@parms{#1}
	\compute@sizes
	\if@scale
                \if@verbose
                        \typeout{psfig: scaling by \@p@scale}
                \fi
                \@ScaleInHundreds{\@p@swidth}
                \@ScaleInHundreds{\@p@sheight}
                \@ScaleInHundreds{\@p@srwidth}
                \@ScaleInHundreds{\@p@srheight}
        \fi
	\ifnum\@p@scost<\@psdraft{
		\if@verbose{
			\typeout{psfig: including \@p@sfile \space }
		}\fi
		\special{ps::[begin] 	\@p@swidth \space \@p@sheight \space
				\@p@sbbllx \space \@p@sbblly \space
				\@p@sbburx \space \@p@sbbury \space
				startTexFig \space }
		\if@angle
			\special {ps:: \@p@sangle \space rotate \space} 
		\fi
		\if@clip{
			\if@verbose{
				\typeout{(clip)}
			}\fi
			\special{ps:: doclip \space }
		}\fi
		\if@prologfile
		    \special{ps: plotfile \@prologfileval \space } \fi
		\special{ps: plotfile \@p@sfile \space }
		\if@postlogfile
		    \special{ps: plotfile \@postlogfileval \space } \fi
		\special{ps::[end] endTexFig \space }
		\vbox to \@p@srheight true sp{
			\hbox to \@p@srwidth true sp{
				\hss
			}
		\vss
		}
	}\else{
		\if@draftbox{		
			\hbox{\fbox{\vbox to \@p@srheight true sp{
			\vss
			\hbox to \@p@srwidth true sp{ \hss \@p@sfile \hss }
			\vss
			}}}
		}\else{
			\vbox to \@p@srheight true sp{
			\vss
			\hbox to \@p@srwidth true sp{\hss}
			\vss
			}
		}\fi

	}\fi
}}
\def\psglobal{\typeout{psfig: PSGLOBAL is OBSOLETE; use psprint -m instead}}
\psfigRestoreAt

\newif\ifpdf
\ifx\pdfoutput\undefined
  \pdffalse
\else
  \pdfoutput=1
  \pdftrue
\fi

\ifpdf
  \usepackage[pdftex]{graphicx}
  \usepackage[pdftex]{color}
  \DeclareGraphicsExtensions{.pdf,.png,.jpg}
\else
  \usepackage[dvips]{graphicx}
  \usepackage[dvips]{color}
  \DeclareGraphicsExtensions{.eps,.epsi,.epsf,.ps}
\fi

\begin{document}

\title{Taking the Initiative with Extempore:\\
Exploring Out-of-Turn Interactions with Websites}

\author{Saverio Perugini$^\dagger$, Mary E. Pinney$^\ddagger$, Naren Ramakrishnan$^\dagger$, Manuel A. P{\'{e}}rez-Qui{\~{n}}ones$^\dagger$\\
$^\dagger$Department of Computer Science\\
$^\ddagger$Department of Industrial and Systems Engineering\\
Virginia Tech, Blacksburg, VA 24061, USA\\
E-mail: \{saverio, mpinney, naren, perez\}@vt.edu\\
\\
Mary Beth Rosson\\
School of Information Sciences and Technology\\
The Pennsylvania State University\\
University Park, PA 16802, USA\\
E-mail: mrosson@ist.psu.edu}
\maketitle

\begin{abstract}
We present the first
study to explore the use of out-of-turn interaction
in websites. Out-of-turn interaction is a technique which empowers the
user to supply unsolicited information while browsing. This
approach helps flexibly bridge any mental mismatch between the user and
the website, in a manner fundamentally different from faceted browsing and
site-specific search tools. We built a user interface~(\mbox{Extempore}) which
accepts out-of-turn input via voice or text; and employed it in a US
congressional website, to determine if users utilize out-of-turn
interaction for information-finding tasks, and their rationale for doing
so. The results indicate that users are adept at discerning when
out-of-turn interaction is necessary in a particular task, and actively
interleaved it with browsing. However, users found cascading information
across information-finding subtasks challenging. Therefore, this work not
only improves our understanding of out-of-turn interaction, but also
suggests further opportunities to enrich browsing experiences for users.
\end{abstract}

\thispagestyle{empty}

\vspace{-0.1in}
\paragraph{Categories and Subject Descriptors:}
H.5.2 [{\bf User Interfaces}]: Interaction Styles;
H.5.4 [{\bf Hypertext/\hskip0ex{}Hypermedia}]: Navigation.

\vspace{-0.1in}

\newpage

\section{Introduction}
It is now well accepted that flexible and contextual web browsing
is imperative for customizing information access. Many solutions
have been proposed---faceted browsing~\cite{Flamenco}, 
personalized search~\cite{pitkow-cacm}, integrating
searching and browsing~\cite{info-scent}, and contextual presentation 
of results~\cite{optimized-search}---all of which
aim to support the user in achieving his or her information seeking goals.
The scope of such research entails 
the development of new interaction techniques~\cite{StrategyHubs,info-scent},
designing interfaces to support these techniques~\cite{Flamenco},
and studying~\cite{ISinEE}/modeling~\cite{interactionScripts}
information-seeking strategies employed by users.
Many of these projects have had qualified success, and one would be
tempted to surmise that all dimensions of research have been thoroughly
explored. In this paper, we identify an additional dimension of information
access that suggests a novel technique for interacting with websites.

\subsection{Setting}
Consider a US Congressional website
organized in a hierarchical
manner, where the site requires the user to progressively make choices of
politician attributes---{\it state} at the first level, {\it branch} at 
the second level, followed by levels for {\it party}, and 
{\it district/seat}---by browsing. Imagine how
a user would pursue the following tasks:

\begin{enumerate}
\itemsep=0pt
\parsep=5pt
\item Find the webpage of the Democratic Representative from District 17 of Florida.
\item Find the webpage of each Democratic Senator.
\end{enumerate}

\begin{figure}
\centering
\begin{tabular}{c}
\includegraphics[width=8.4cm]{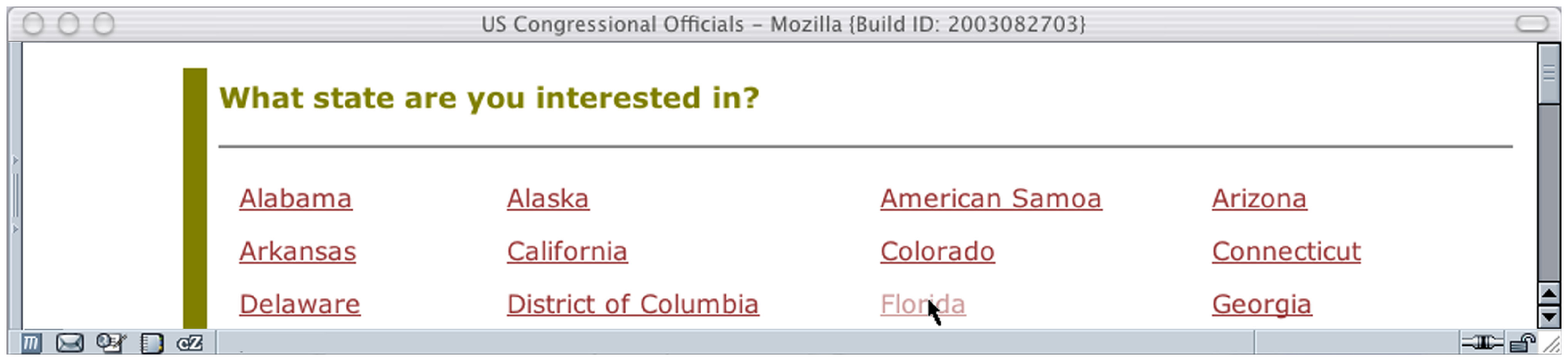}\\
$\mathbf{\Downarrow}$\\
\includegraphics[width=8.4cm]{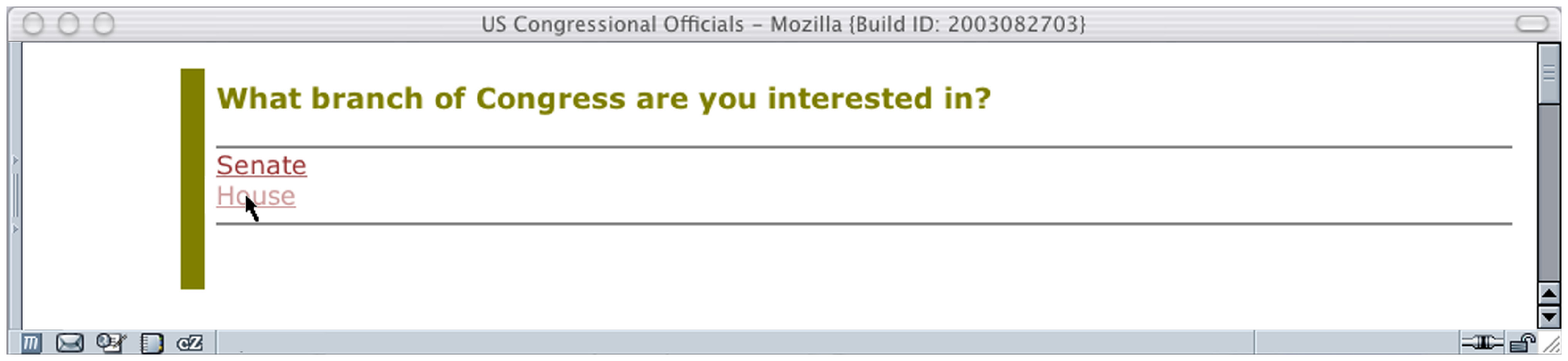}\\
$\mathbf{\Downarrow}$\\
\includegraphics[width=8.4cm]{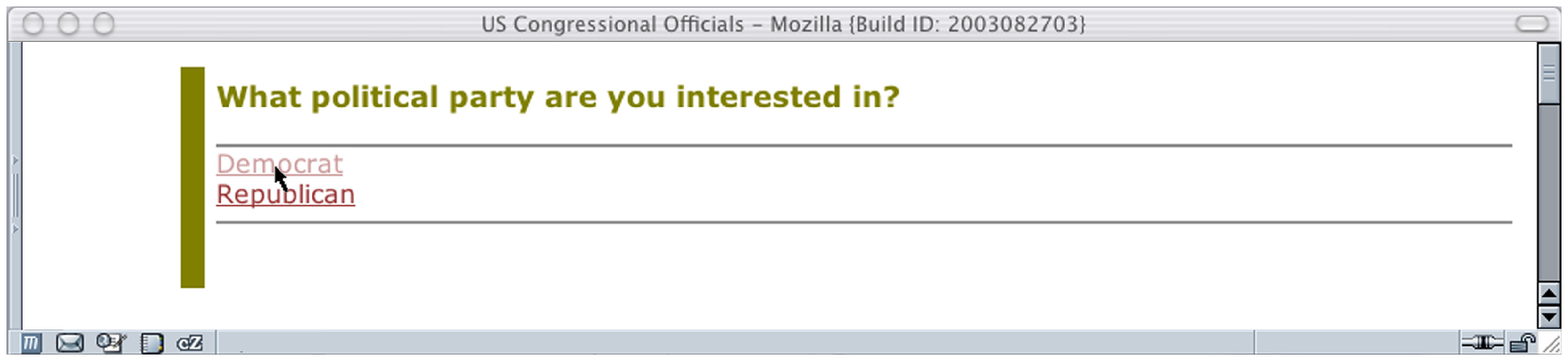}\\
$\mathbf{\Downarrow}$\\
\end{tabular}
\caption{In-turn interaction with a Congressional site.}
\label{pvs-hier}
\end{figure}

The first task can be satisfied by typical drill-down browsing because
it involves supplying {\it in-turn}, or responsive, information at each
level (see Fig.~\ref{pvs-hier}). By in-turn, we mean that the user need only click on presented
hyperlinks~(click `Florida' first, `House' next, and so on). Each
click communicates {\it partial information} about the desired politician.
Achieving the second task by communicating only in-turn information
would require a painful series of drill-downs and roll-ups, in order
to identify the states that have at least one Democratic Senator, and to aggregate
the results.
While the user has partial information about the desired
politicians, s/he is unable to communicate it by in-turn means. 

The key observation here is that flexibility of information access
will be enriched by increasing the means for supplying partial information.
Ideally, the user, having seen that s/he does not have the partial
information requested at the top level~(i.e.,~state),  would have liked to
supply the information that s/he {\it does} have, namely that of party and
branch of Congress. 

\subsection{Solution Approach}
{\it Out-of-turn interaction} is our solution to support
flexible communication of partial information not
currently requested by the system.
Hence such information is unsolicited but presumably relevant to the
information-seeking task. Out-of-turn interaction is thus unintrusive,
optional, and can be introduced at multiple points in a browsing session,
at the user's discretion.
One possible means to support it is to allow the speaking of utterances
into the browser. 

\begin{figure}
\centering
\begin{tabular}{c}
\includegraphics[width=8.4cm]{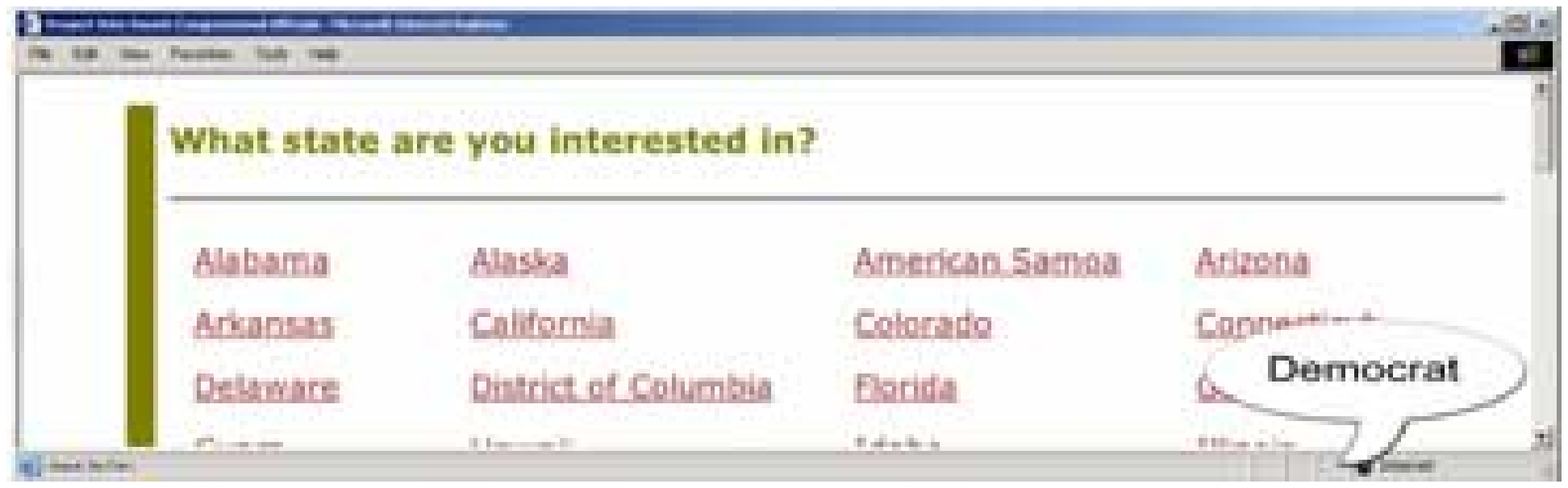}\\
$\mathbf{\Downarrow}$\\
\includegraphics[width=8.4cm]{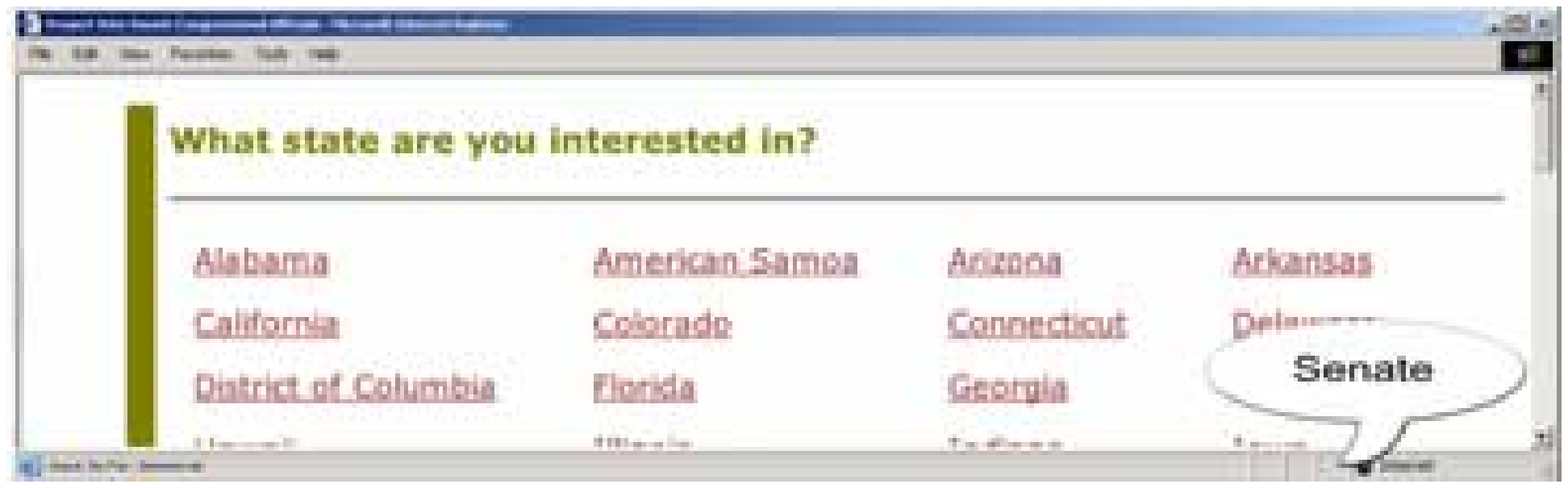}\\
$\mathbf{\Downarrow}$\\
\includegraphics[width=8.4cm]{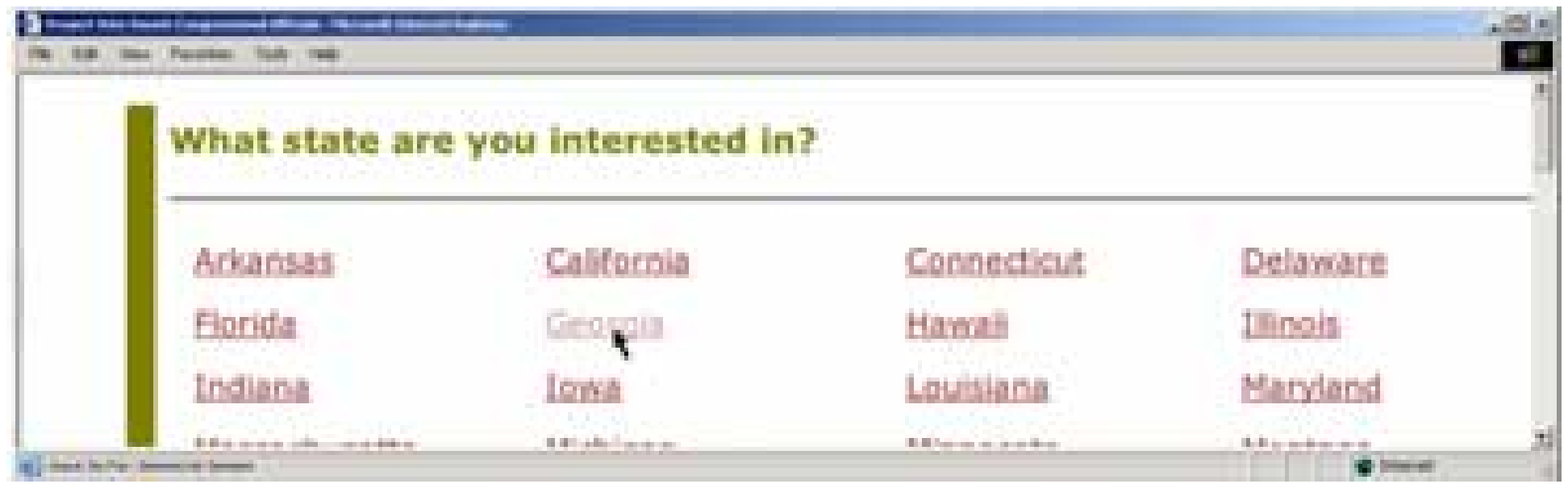}\\
$\mathbf{\Downarrow}$\\
\includegraphics[width=8.4cm]{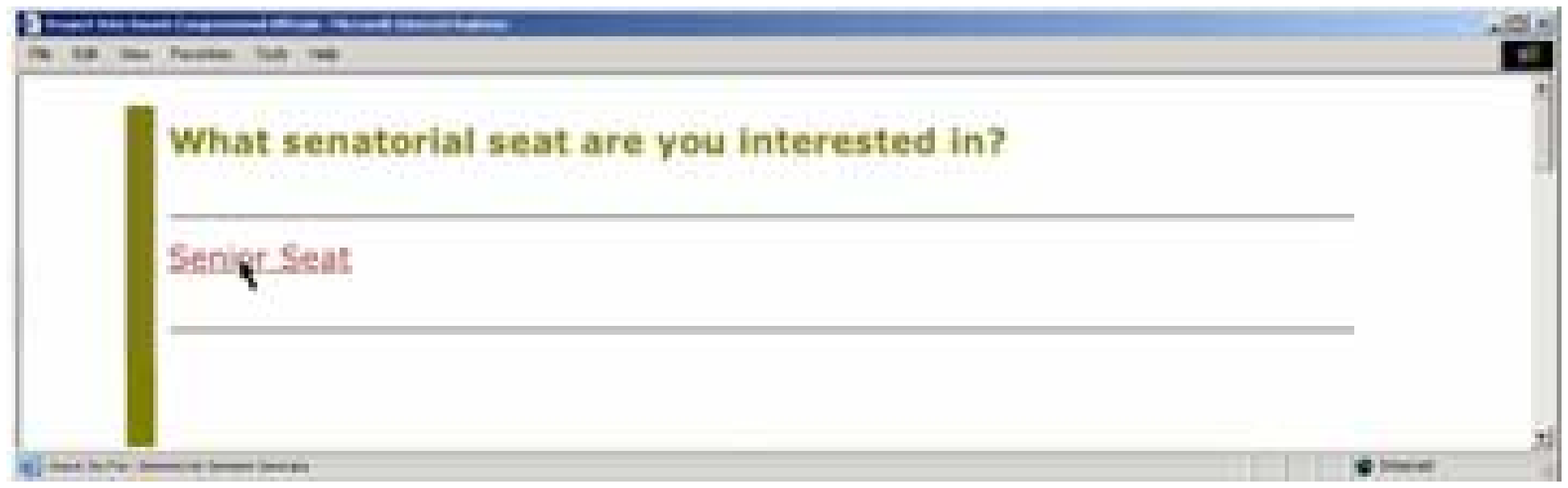}\\
$\mathbf{\Downarrow}$\\
\includegraphics[width=8.4cm]{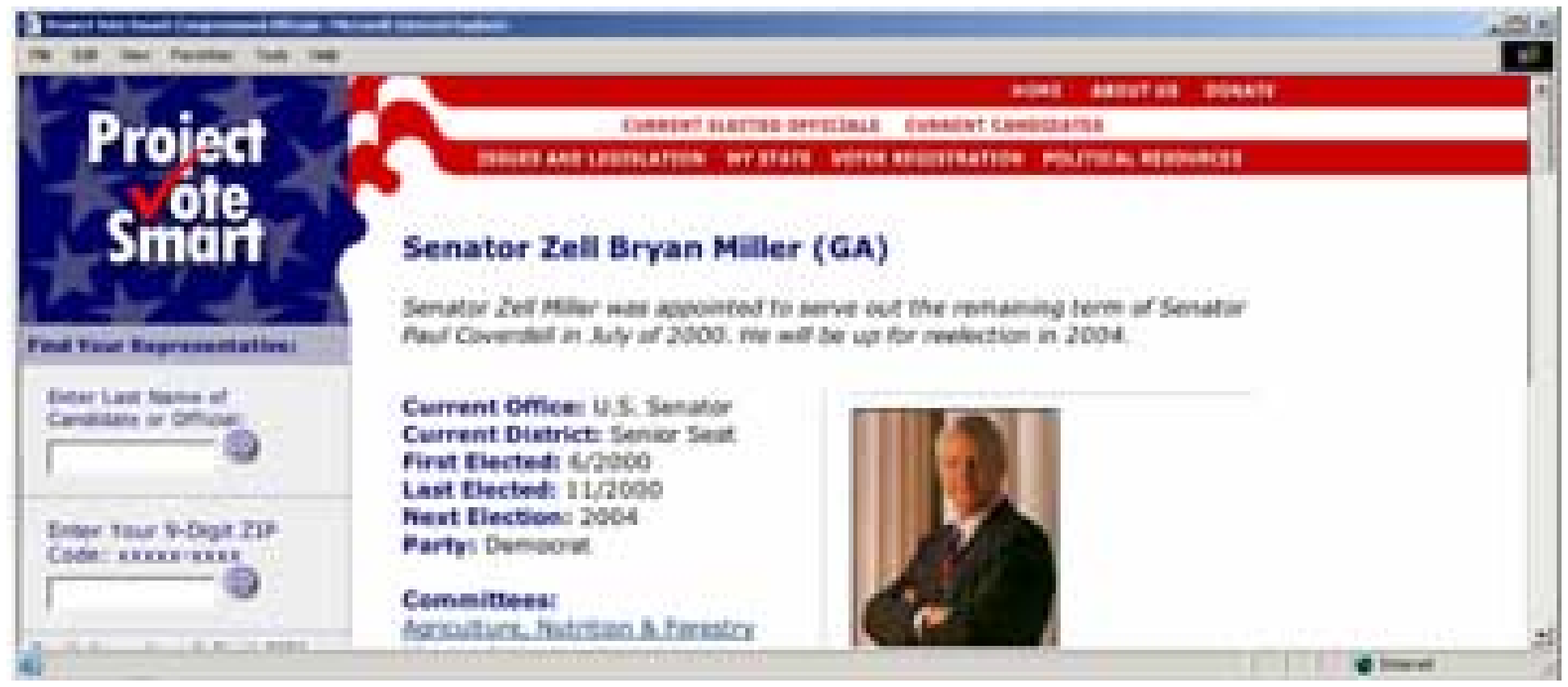}
\end{tabular}
\centering
\caption{A web session illustrating the use of out-of-turn
interaction in a US congressional site.
This progression of interactions shows how the~(Democrat, Senate, Georgia,
Senior)~interaction sequence, which is
indescribable by browsing, may be realized. In steps 1 and 2,
`Democrat' and `Senate' are spoken out-of-turn~(resp.) when 
the systems solicits for state. In step 3, the user clicks `Georgia'
as the state (an in-turn input). The screen at Step 4 shows that only the 
Senior Senator from Georgia is a Democrat, and leads the user to his homepage.}
\label{toolbar-dialog}
\end{figure}

Figure~\ref{toolbar-dialog} describes using out-of-turn interaction to
achieve Task 2 above. At the top level of the site, the user is
unable to make a choice of state, because s/he is looking for states
that have Democratic Senators. S/he thus speaks `Democrat' out-of-turn,
causing some states to be pruned out~(e.g.,~Alaska). At the second
step, the site again solicits state information because this aspect
has not yet been communicated by the user. The user speaks `Senate'
out-of-turn, causing further pruning~(e.g.,~of~American Samoa),
and retaining only regions that have 
Democratic Senators. At this point, the goal has been
achieved~(the user notices 31 states satisfying the criteria), and 
s/he proceeds to browse through
the remaining hyperlinks. Notice that these are contextually relevant
to the partial information supplied thus far, so that when `Georgia'
is clicked, there is only one choice of seat~(Senior) implying that
the other Senatorial seat is not occupied by a Democrat.

\section{Out-of-turn Interaction}
What does it mean to interact out-of-turn? One interpretation is that,
when the user speaks `Democrat', s/he is desiring to experience 
an interaction sequence through the site containing `Democrat.' The implicit
assumption in the current implementation is that what is spoken
is a link label (or variation thereof) nested deeper in the site, 
and hence an in-vocabulary utterance\footnote{In other implementations,
we might conduct a more elaborate modeling of the vocabulary.}. Therefore, out-of-turn interaction 
is merely a mechanism to address alternate aspects of the given activity,
while postponing the specification of currently solicited aspects. 

Why would users interact out-of-turn? There are several reasons. First, 
what the site is requesting from the user may actually be what the user
is seeking in the first place! For example,
in Figure~\ref{toolbar-dialog}, the site is soliciting state but
the user is
looking for states with a certain property. Second, being able to speak
out-of-turn in an otherwise hardwired site permits the realization of
interaction sequences not describable by browsing. This means that we can
support all permutations of specifying politician attributes, without
explicitly enumerating in-turn choices. The above example
would require 4!=24 faceted browsing 
classifications
to support all
tasks~(i.e.,~browse by party-state-branch-district, by 
state-party-branch-district, and so on). Third, the incorporation of
out-of-turn information does not curb the interaction~(i.e.,~the
levelwise organization is preserved), but rather situates future interactions 
in the context of past ones. 
More fundamentally, out-of-turn 
interaction is a novel way to flexibly bridge any mental
mismatch
between the user and the website, without anticipating when the mismatch 
might happen.

\subsection{Related Research}
\begin{figure}
\centering
\includegraphics[width=7.5cm]{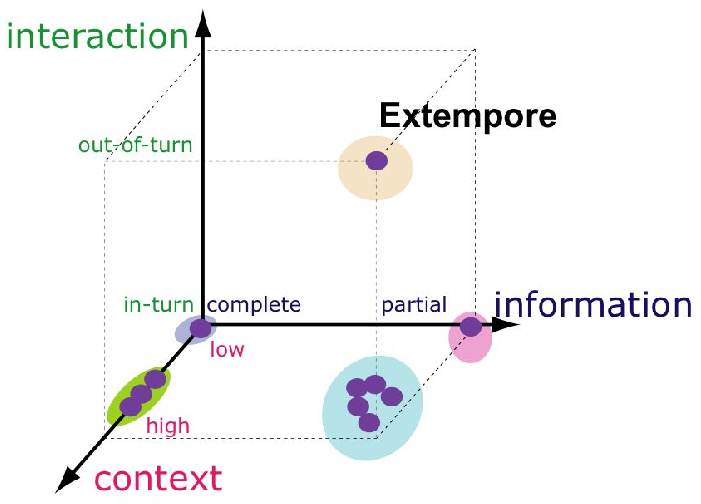} 
\caption{Three dimensional space showcasing related research. Each of the shaded
clusters denotes a concerted group of projects discussed in the main text.}
\label{mccricks}
\end{figure}

To better understand the merits of out-of-turn interaction, we showcase
related research in a three-dimensional space (see Figure~\ref{mccricks})
involving:~(i)~the nature of information exploited, (ii)~the level of context
supported, and~(iii) the interaction technique.

The first axis distinguishes between the specification of partial
vs. complete information. Supporting only the specification of 
complete information means that interaction is viewed as a 
one-shot activity; supporting specification of partial information 
implies that information-seeking is to be conducted over multiple
steps. Since the complete information approach is more restrictive
than the partial information approach, it is situated toward the
origin. The second axis makes a distinction of whether input or results~(or both)
are contextually qualified in some manner. Our contribution to this
space is the third dimension of whether interaction occurs by
in-turn or out-of-turn means.

Search engines~(e.g.,~Google) are characterized
by specification of complete information~(in this case,
the query), because the interaction is terminated by returning 
a flat list of results. Such a low-context, complete information 
approach is denoted by the origin in Fig.~\ref{mccricks}.
Browsing, on the other hand, involves the incremental specification of partial 
information~(right of origin in Fig.~\ref{mccricks}). 

When we take context into account, two further clusters of projects
emerge in the in-turn plane spanning the~(information~$\times$~context)
axes. When only complete information
is supported, results presentation provide the major opportunity for
exhibiting context~(front left corner of Figure~\ref{mccricks}). This 
is seen in site-specific search tools~(e.g.,~at Amazon.com), in 
the contextual search of Dumais et al.~\cite{optimized-search},
and the personalized search strategies of Pitkow et al.~\cite{pitkow-cacm}. The more
dense cluster~(front right of Figure~\ref{mccricks})
forms in the partial information region. 
These are projects that support contextual information access by 
providing either greater input flexibility or adaptable display
of results over the course of an interaction, or both.
Faceted~(flat or hierarchical) organizations~\cite{Flamenco,polyarchiesCHI}, 
Dynamic Taxonomies~\cite{DynamicTax},
Strategy Hubs~\cite{StrategyHubs}, adaptive hypermedia~\cite{adaptiveHypermediaSurveyJoP2},
and ScentTrails~\cite{info-scent}
are examples. We discuss these further.

Sites and systems exposing faceted browsing~(e.g.,~epicurious.com)
support multiple classifications by 
providing enumerated in-turn choices. This often leads to cumbersome site designs
and a mushrooming of possible choices at each step.
The Dynamic Taxonomies project provides in-turn operators for pruning
information hierarchies, while
Strategy Hubs enumerates templates for prolonged and detailed
information-seeking tasks, again in-turn. 
The adaptive hypermedia projects employ user models (e.g., of
past browsing behavior) to tailor the presentation of hyperlinks.
ScentTrails argues that browsing may not be focused enough and that searching
loses context, and aims to combine them in a single
framework. However, its use of searching always precedes browsing and
therefore limits the richness of supportable interactions. Out-of-turn
interaction aims to provide precisely this combination of focused input
and exploratory browsing in a single, flexible, framework. 

Our work can be viewed as complementary to these efforts in that it
{\it lifts} the nature of interaction from in-turn to out-of-turn
means~(top of Figure~\ref{mccricks}).
For instance, the example session shown in Figure~\ref{toolbar-dialog}
can be viewed as a lifted version of traditional browsing, 
yielding a~(high context, out-of-turn) technique that exploits
partial information. While out-of-turn interaction can be studied in many
settings, this paper only discusses its use in conjunction with 
browsing of levelwise, hierarchically organized sites.

Out-of-turn interaction, especially of the unsolicited reporting nature,
has been recognized as a simple form of mixed-initiative 
interaction~\cite{MII-UR}.
Interleaving out-of-turn responses with in-turn clicks can be viewed as
conversational shifts of initiative between the user and the website.

\subsection{Extempore}
\begin{figure}
\centering
\includegraphics[width=8.4cm]{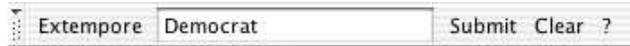}
\caption{The Extempore \mbox{out-of-turn} interaction toolbar. This
interface is embedded into a traditional web browser to augment hyperlink
interaction.
User has supplied Democrat presumably out-of-turn.}
\label{toolbar}
\end{figure}

We have built a user interface, called {\it Extempore}, that accepts
out-of-turn input either via voice or text. 
The voice version was implemented
using SALT 1.1~(\cite{SALT};
a standard that augments HTML with tags for speech 
input/output) and SRGS~(Speech Recognition Grammar Specification), for 
use with Internet Explorer 6.0. The text version is a toolbar embedded
into the Mozilla/\hskip0ex{}Netscape web browser~(v1.4)
and was implemented using XUL~(see
Figure~\ref{toolbar})\footnote{Currently there is no SALT plugin for
Mozilla~(and likewise with XUL and IE). Due to these technological
constraints, we do not support both interfaces of Extempore in the
same implementation.}.
It is important to note that Extempore is embedded in the web browser,
and not the site's webpages. It is also
not a site-specific search tool
that returns a flat list of results~(akin to the Google toolbar). 
Further, while search engines index webpages, Extempore rather relies
on an internal representation of the website and, when out-of-turn input
is supplied, uses transformation techniques to stage the
interaction, pruning the website accordingly. The details of 
the underlying software transformations are beyond the scope of 
this work; see, for example, Ricca and Tonella~\cite{webAppSlicing}, 
and Perugini and Ramakrishnan~\cite{MII-ITPro} 
for ideas on transformation techniques. Extempore can be used for
out-of-turn interaction in many web sites, given a representation of
the site's structure, e.g., in XML.

\section{Exploratory Study}
Extempore is fundamentally different from existing approaches
to customize browsing experiences; therefore we conducted a study
that exposes users to out-of-turn interaction, to determine if
they utilize it in information-finding tasks, and their rationale for
doing so. The main component of the study entailed asking participants
to perform eight specific information-finding tasks
in the Project Vote Smart~(PVS) website~(http://www.vote-smart.org)\footnote{At
the time this study was conducted,
PVS employed a hardwired organization akin to Figure~\ref{pvs-hier};
the site has been recently restructured into a flat faceted classification.}.
Rationale was gathered through think-aloud and retrospective protocols.

\subsection{Goals}
The goal of the experiment was to study usage patterns for out-of-turn 
interaction, not to evaluate the interfaces used to realize it, or
to compare out-of-turn interaction with other interaction techniques.

\subsection{Participants}
We collected data from 24 participants in the analysis; all were students
with an average age of 21, and a majority were undergraduates
in computer science. Some of the participants were recruited from
a HCI course, and were compensated with extra-credit from the instructor.
Since a component of this experiment involved
voice recognition software, we primarily recruited native speakers of
English.  Average participant computer and web familiarity and use
was 4.75 or greater on a 5-point Likert scale. Average participant 
familiarity with voice recognition software was 1.46, and
mean familiarity with the 
structure of the US Congress was 2.83; no user had visited
the PVS website prior to the experiment.

\subsection{Tasks}
The eight tasks were carefully formulated to generate a diverse set
of interaction choices:
\begin{description}
\itemsep=0pt
\parsep=5pt
   \item[A.] Find the webpage of the Junior Senator from New York.
   \item[B.] Find the webpage of the Democratic Representative from District 17 of Florida.
   \item[C.] Find the webpage of the Republican Junior Senator from Oregon.
   \item[D.] Find the webpage of the Democratic member of the House in Rhode Island
serving district 2.
   \item[E.] Find the states which have at least one Democratic Senator.
   \item[F.] Find the states which have twenty or more congressional districts.
   \item[G.] Find the states which have at least one Republican member of
the House.
   \item[H.] Find the political party of the Senior Senator representing the
only state which has congresspeople from the Independent party.
\end{description}

\begin{figure}
\centering
\includegraphics[width=8.4cm]{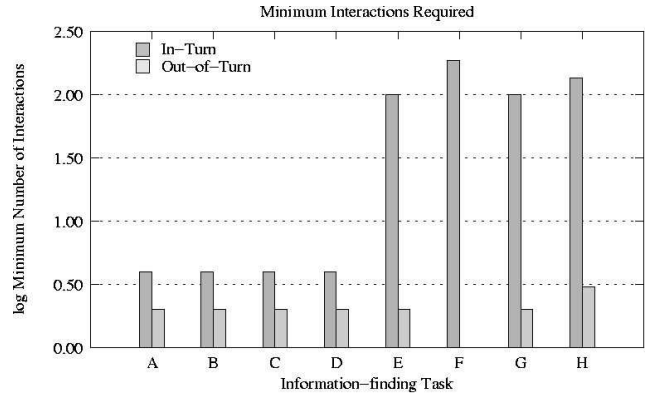}
\caption{Minimum number of interactions~(log10 scale) required to successfully satisfy
each information-finding task using in-turn~(dark) and out-of-turn~(light)
interaction.  Note that Task F can be completed with just one 
out-of-turn interaction, so its entry in the graph shows zero.}
\label{minimum}
\end{figure}

We refer to tasks A, B, C, and D as {\it non-oriented} tasks,
in that they
can be performed as easily
by employing solely in-turn interaction~(i.e.,~in this case,
hyperlinks),~solely out-of-turn interaction~(Extempore), or using a 
mixture of both.  
Out-of-turn interaction
does not appear to be worthwhile with respect to
these tasks because the effort required to perform them with out-of-turn
interaction is commensurate with that of in-turn interaction.
Tasks E, F, G, and H are {\it out-of-turn-oriented},
because they are difficult to perform with only in-turn interaction.
Formally, we say an information-seeking task is out-of-turn-oriented 
if the minimum number of browsing
interactions required to successfully complete 
it exceeds the maximum depth of the targeted website; 
otherwise it is non-oriented. 

The maximum depth of the PVS site 
is four and Figure~\ref{minimum} illustrates the
minimum number of interactions required per task. 
In calculating this minimum number, we assumed that the user
can supply at most one aspect at each step~(in-turn or out-of-turn),
and discounted back button clicks~(happens when employing only
in-turn interaction for an out-of-turn-oriented task). Notice
also that some tasks, namely the non-oriented ones,
cannot be performed by purely a sequence
of out-of-turn interactions; a terminal in-turn input is
often necessary and these are discounted as well. For instance,
try solving task A using purely out-of-turn inputs.

\subsection{Design}
The study was designed as a within-subject experiment.
Task was the independent variable and
the interaction observed~(in-turn vs. out-of-turn)
was the dependent variable.
Participants were given both the toolbar and voice interface of
Extempore; and performed four tasks with each~(two non-oriented and two
out-of-turn-oriented). We designed the experiment with the 
provision for interfaces in two different modalities, to more
naturally assess the use of out-of-turn interaction
independent of a particular interface for it.
Each participant performed the eight tasks in an order pre-determined
by a latin square to control for unmeasured factors.
In addition, the specific interface to be used~(toolbar or voice)
for a~(task,~participant) pair was determined
{\it a priori} by complete counterbalancing within each task category.
Thus, for each task, half of the twenty-four
participants were given the toolbar
interface and half the voice interface.
The participants were free to utilize any strategy
to complete the information-finding tasks, given
Extempore and the available hyperlinks; they were given 
unlimited time to complete each task.

\subsection{Configuring Extempore}
A vocabulary for the PVS site was created by collecting all link labels,
synonyms~(e.g.,~`Representatives' for `House'), and alternate forms of common
utterances~(e.g.,~`Senate', `Senator', `Senators'). 
Both the toolbar and voice version of Extempore supported this vocabulary,
with the toolbar supporting abbreviations~(e.g.,~CA for California), in 
addition. To keep users abreast of partial information supplied thus
far~(either by browsing or via Extempore), we continually updated
an `Input so far:' label in the browser status bar~(see Figure~\ref{toolbar-dialog}).
We also included a provision for the user
to inquire about what partial information is left unspecified at any step.
Access to this feature is provided through a `What May I Say?' button~(labeled
with a `?' in Figure~\ref{toolbar}) or utterance.

The semantics of out-of-turn interaction in information hierarchies
required some practical implementation decisions. For instance, when
the user speaks `Junior seat,' the specification of `Senate' can be 
automatically inferred by functional dependency. 
Another form of 
such `utterance expansion' occurs in response to single-valued options.
For instance, in Figure~\ref{toolbar-dialog}, one can argue that the
choice of seat at the last step is really unnecessary, as there is
only one option left~(Senior). When only one path remains among
the available options, we vertically collapse them and directly 
present the leaf page. This feature was not illustrated in 
Figure~\ref{toolbar-dialog} for ease of presentation, but we implemented it
in our study. Notice, however, that no information is lost
during such collapsing, since terminal pages in PVS identify all pertinent
attributes of politicians.

\subsection{Equipment, Training, and Procedures}

\subsubsection{Equipment}
Participants performed the tasks on an Extempore-enabled
Pentium III workstation, connected to a 17" monitor set at
2560$\times$1024 resolution in 34-bit true color, running Windows 2000.
We recorded a video
of each participant performing the information-finding
tasks using the Camtasia screen and
audio capture software. The resulting capture
was used to aid participant recollection during the retrospective
verbal protocol as well as in subsequent analysis~(e.g.,~think-aloud).
The Audacity audio recording application was used during
the retrospective portion of the experiment to capture participant
explanations. Data from the pre-questionnaire~(demographics, computer
familiarity) and post-questionnaires~(rationale) was recorded on
paper. Finally at the end of the entire experiment
we transcribed and collated the data gathered from all sources to
construct a complete record of each participant session,
including interaction sequences followed per task. Each participant
session lasted approximately 90 minutes.

\subsubsection{Training}
Prior to revealing the information-seeking tasks, we gave
users specific training on~(i)~the PVS website, including
levels of classification, and
interacting with it via hyperlinks;~(ii)~interacting with
PVS using Extempore~(both toolbar and voice interfaces);
and~(iii)~interleaving hyperlink clicks with 
commissions via Extempore.  Users were provided a card
summarizing the vocabulary that Extempore can understand,
as well as explanations of political terms and their functional
dependencies. This card was available for their use during the
entire session, not just training. We did not use terms such as
`in-turn' or `out-of-turn' during training or elsewhere in
the study. This is to prevent biasing of participants
toward any intended benefits of Extempore, and also to help
them conceptualize its functionality on their own.
In other words, we simply trained users on how
to employ the available interfaces~(hyperlink and Extempore)
for information seeking. After some self-directed exploration,
users were given a short test consisting of four practice
tasks~(two with toolbar and two with voice).

\subsubsection{Procedures}
After the users completed the training tasks, we administered
the actual test involving tasks A--H above, and employed
both concurrent~(think-aloud) and retrospective protocols to
elucidate rationale. A structured interview, including a
post-questionnaire, was conducted to gather additional
feedback. The entire experiment
generated~(24$\times$8~=)~192~(participant,~task) interaction sequences. 

These sequences were then analyzed for frequencies of usage of
in-turn vs. out-of-turn interaction.
For purposes of this study, we defined
an in-turn interaction as a hyperlink click or the communication of
in-turn partial information to the website via Extempore. Notice
that just saying 'Connecticut' will not qualify as an out-of-turn
interaction, if the same choice was currently available as a hyperlink.
Similarly, we defined an out-of-turn interaction to be the submission
of one aspect of unsolicited partial information to the site.
Supplying more than one aspect of partial information to the
site out-of-turn~(e.g.,~saying `Democratic Senators')
corresponds to multiple out-of-turn interactions.

Notice that a user may supply
in-turn and out-of-turn information to the website simultaneously
via Extempore. For instance, in the top-level page in Figure~\ref{toolbar-dialog},
the user might say `House, Florida, District 17, Democrat,' all
at the outset. Observe that a permutation of this utterance exists---`Florida,
House, Democrat, District 17'---that, if conducted incrementally, could imply 
a purely in-turn interaction. Such an interaction is thus viewed as having
four in-turn inputs. On the other hand, consider a user who says
`New York, Democrat' at the outset. There is no permutation with
respect to the
PVS site that permits viewing this utterance as comprising of purely
in-turn input, and hence, it is classified as one in-turn
input~(`New York'), followed by an out-of-turn input~(`Democrat').
This policy of counting does not favor~(and actually deprecates)
out-of-turn interaction.

Some users, after completing a given task via out-of-turn
interaction, verified part of their results via in-turn interactions.
This was confirmed through their retrospective feedback, and
such in-turn interactions were discounted in the analysis.

\section{Results}
Of the 192 recorded interaction sequences, 
177 of them involved the
successful completion of the task by the participant. We analyze these
177 sequences first, followed by the remaining 15
sequences~(which were all generated in response to 
out-of-turn-oriented tasks).

\subsection{General Usage Patterns}
Results indicate a high frequency of usage for out-of-turn interaction.
94.4\% of the 177 sequences contained at least one out-of-turn
interaction. In addition, every participant used out-of-turn interaction
for at least 70\% of the tasks, with 16 people using it in all tasks.
Conversely, every task was performed with out-of-turn interaction by
at least 80\% of the participants, with 4 tasks enjoying
out-of-turn interaction by all participants. These results are encouraging
because Extempore usage is optional and not prompted by any indicator
on a webpage. 
Participants successfully
completed the given tasks
irrespective of the presented interface~(voice or toolbar).

\subsection{Classifying Interaction Sequences}
The 177 interaction sequences were classified into five categories
denoted by:~(i)~I, (ii)~O, (iii)~IO, (iv)~OI, and~(v)~M. The I and O
categories denote sequences comprised of purely in-turn, or out-of-turn
inputs, respectively. In IO sequences all in-turn inputs precede out-of-turn
inputs~(analogously, for OI). M sequences~(`mixed') are those which
do not fall in the above categories. For instance, the interaction shown
in Fig.~\ref{pvs-hier} would be classified under I, and that in
Fig.~\ref{toolbar-dialog} is in OI.
We posit that this classification 
provides insight into users' information-seeking strategies, and can be
related to the nature of the information-finding task.

\begin{figure}
\centering
\includegraphics[width=8.4cm]{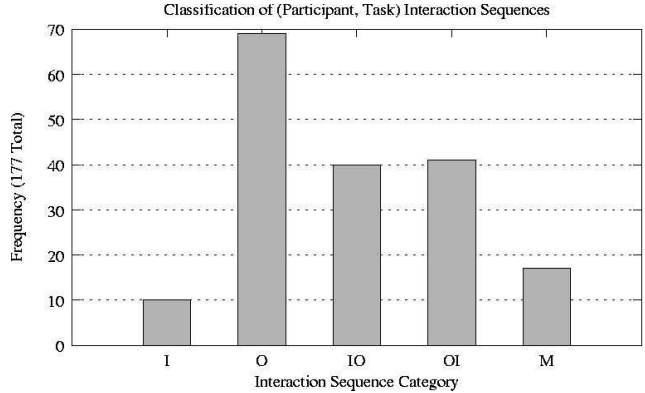}
\caption{Classification of 177 (participant, task)
interaction sequences.} 
\label{sequence-cats}
\end{figure}

Figure~\ref{sequence-cats} shows the distribution of the 177 sequences into
the five classes, and Table~\ref{twobox} depicts a breakdown by
both task orientation and classes. Notice that O, OI, IO, and mixed 
classes have
been grouped in Table~\ref{twobox} to distinguish them from pure
browsing interactions~(I).  

As Figure~\ref{sequence-cats} shows, 10 of the 177 sequences fall in the I
class, i.e., these are {\it browsing} sequences. 
As Table~\ref{twobox}~(lower left) shows, all of the 10 browsing
sequences were generated in response to non-oriented tasks, revealing
that a 100\%~(81/81) of the sequences for out-of-turn-oriented
tasks involved out-of-turn interaction.
Therefore,
\begin{itemize}
\itemsep=0pt
\parsep=5pt
\item users never attempted to achieve an out-of-turn oriented task via browsing;
or in other words,
\item users always employed out-of-turn interaction when presented with an 
out-of-turn-oriented task.  
\end{itemize} 

{\bf This is notable because it confirms that users are adept at discerning 
when out-of-turn interaction is necessary.}

\begin{table}
\centering
\begin{tabular}{l|r|r|r}
   		      & {\bf I} & {\bf {\{O,IO,OI,M\}}} & {\bf total}\\ \hline
{\bf non-oriented}    	& 10      & 86  & 96 \\ \hline
{\bf out-of-turn-oriented} & 0    & 81  & 81 \\ \hline
{\bf total}           & 10      & 167 & 177 \\ 
\end{tabular}
\caption{Breakdown of 177 interaction sequences in various categories. The total number
of interaction sequences for out-of-turn oriented tasks is 15 less than that for
non-oriented tasks; these were the sequences where the participant did not
complete the task successfully.}
\label{twobox}
\end{table}

\subsection{Detailed Analysis of Interaction Classes}
Let us now study the interactions in classes O, OI, IO, and M.
The 69 pure out-of-turn sequences~(O) were observed only in out-of-turn-oriented
tasks E, F, and G, and was used by all the 24 participants. This
clustering of the O sequences around three tasks shows that,
whenever participants completed these tasks, they did so in the
shortest manner possible. Refer again to Figure~\ref{minimum} for the
sharp contrast in the length of the minimum out-of-turn sequence from
the minimum in-turn sequence, for these tasks.

Classes IO, OI, and M contain the sequences exhibiting rich interaction
strategies. Classes IO and OI were observed in near-equal numbers,
and primarily in the non-oriented tasks~(A, B, C, and D)
with the exception of OI, which was also seen in task H.
No particular clustering was observed with respect to participants. The 17 class M
interactions exhibited only two types of patterns---14 with an OIO form,
and 3 with an IOI form. Furthermore, like OI, these 17 mixed
interactions also involved only the non-oriented tasks~(A, B, C, D) and task H.
It is interesting that we are observing OIO and IOI sequences, even in
a site with only four levels. Once again, no specific clustering
was observed around participants.

To see if these classes correspond to specific information-seeking
strategies, we plotted curves depicting the
progressive narrowing down to a desired congressional official, as 
a function of interaction steps. All curves begin at the~(0,~540) point
because the PVS site indexes all 540 congressional officials. With 
each interaction, this number is gradually reduced until 
the user arrives at the desired set of officials. However, we were unable to 
observe major correlations between curve slopes and strategies;
this is because in the PVS site, the slope is primarily dependent 
on the nature of the task, not the strategy. For instance
if a task involved a state like `Rhode Island,' even an in-turn input
of this state information will cause greater pruning than most
out-of-turn inputs. To qualify interaction classes better, we must
study out-of-turn interaction in more sites.

\subsection{Cascading Information across Subtasks}

\begin{figure}
\includegraphics[height=3.75cm,width=6.5in]{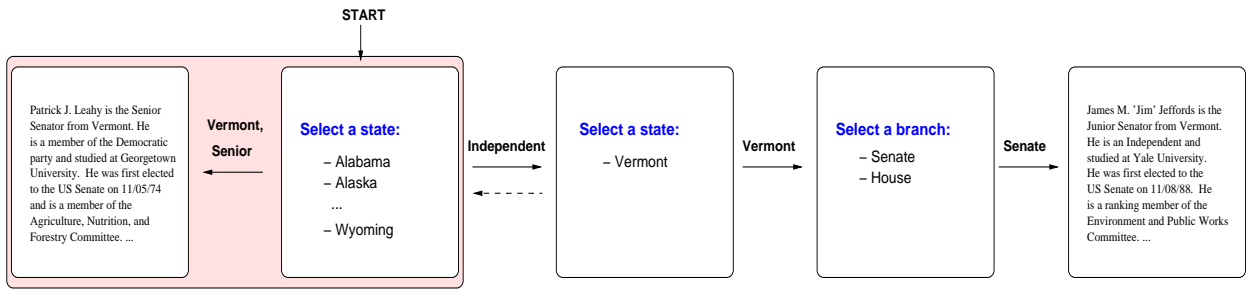}\\
\caption{Task H: the user is expected to first find `Vermont' in one
Interaction~(third window from left) and use it as input
in another interaction~(shaded area) to find the
party of the Senior Senator from that state. The two windows on
the right depict unnecessary and irrelevant interactions for this task.}
\label{confusing}
\end{figure}

Recall that 15 interaction sequences led to incorrect answers;
interestingly 12 of these 15 were generated in response to
Task H. Notice that Task H is challenging, because it involves 
two subtasks and cascading
information found in one into the other. The user is expected to first
find the only state having Independent congressional officials~(Vermont),
and then find the political party of the Senior Senator from that
state~(Democrat). In other words, this task requires procedural, not 
just declarative, knowledge~(a distinction motivated in the
Strategy Hubs project~\cite{StrategyHubs}).

Most people were adept at finding that Vermont was the desired
state~(e.g.,~by saying `Independent' at the outset), but did not realize 
that the task cannot be completed by {\it continuing} that interaction.
As Figure~\ref{confusing} shows, clicking on the only available
state link~(`Vermont') now presents a choice of House vs. Senate.
Clicking on Senate takes the user to the webpage of Jim Jeffords, who
is the Junior Senator from Vermont, not the Senior Senator! 

Some users immediately realized the problem, as identified in their
retrospective interviews, e.g.:
\begin{quote}
``This question was tricky. Cause it was, I was like wait, if he's
Independent then his party is Independent \ldots at first~[I thought]
it was the Senior Senator who was Independent \ldots and I got this 
guy's webpage, and then I saw that he was the Junior \ldots So then 
I eventually went back to Vermont and got the~[Senior] guy.''
\end{quote}
Only 12~(50\%) of the participants successfully completed this task.
This result demonstrates that cascading information across subtasks is
challenging. It was clear that all users wanted to continue the
interaction, but some failed to realize that out-of-turn interaction
as presented here is merely a pruning operator, and not 
constructive. Investigating the incorporation of constructive operators
such as rollup/\hskip0ex{}expansion is thus a worthwhile direction of 
future research.

\subsection{Rationale and Qualitative Observations}
Studying users' rationale revealed their reasons for interacting out-of-turn:
\begin{quote}
``I can jump through all the levels \ldots.''

``Initially I thought I would prefer the hyperlinks \ldots after
reading the questions, it became apparent that the toolbar
and voice interface would simplify the task.''

``\ldots when you wanted to know all the states for the Republicans,
then you would have to click on every single link. It would just
get annoying after a while. You'd just give up I think. There'd be
no way.''

``I guess I would have had to \ldots wow, check every state.''
\end{quote}
demonstrated understanding of how Extempore works~(e.g.,~input expansion):
\begin{quote}
``Its the easiest way cause there is only one Representative from District
17 in Florida and it takes you straight to the page.''

``If you click on the state then you get choices of House and
whatever, but if you type in district 2 and it just goes right there.''
\end{quote}
presented advantages and judgments:
\begin{quote}
``\ldots allowed multiple pieces of information to be input at one time.''

``As much surfing as I do, it sort of makes me wish I had those
options sometimes ya know instead of going to search engines and fooling
around \ldots having to come up with different search criteria \ldots.''
\end{quote}
and also brought out frustrations:
\begin{quote}
``The voice interface feels a little awkward since I am not used
to talking to myself \ldots.''

``I don't always trust the results,~[so I went back] confirming that
they are all republican.''
\end{quote}

Many users learned that out-of-turn interaction is best suited when 
they have a specific goal in mind, and not meant for exploratory
information-seeking~(as is browsing). For instance,
\begin{quote}
``if I wanted to go the whole way down to a specific person, I would use~[Extempore],
but if I was just looking around, I would use the links.''

``[Extempore] is good when you know the site and know you have to
go several layers deep. Links~[are good] when you don't know the layout
or don't know exactly what you want.''
\end{quote}

\section{Discussion}
Extempore enables a novel approach to interact with websites. Users with out-of-turn
partial input can employ Extempore to enhance their browsing experiences.
Thus, out-of-turn interaction is intended to complement browsing, and not replace it.
For designers, Extempore augments their sites with capabilities for personalized
interaction, without hardwiring in-turn mechanisms~(as is commonly done).
In addition, since usage of Extempore is optional, it preserves any existing
modes of information-seeking. 

There are significant lessons brought out by our study, which we only briefly
mention here. This work validates our view of web interaction as
a flexible dialog and shows that users actively interleaved 
out-of-turn interaction with browsing. Importantly, users
were proficient at determining when out-of-turn interaction is called for.
Studying the rationale and usage patterns has generated
a body of knowledge that can be used, among other purposes, for introducing
out-of-turn interaction in new settings and to new participants. Furthermore,
we have seen that it is easy to target out-of-turn interaction in domains
where tasks involve combinations of focused and exploratory behavior.
Recall also that dialogs with purely declarative specifications are readily
supported; others such as Task H will require further study.

Out-of-turn interaction is most effective when users have a basic understanding
of the application domain and know what aspects are addressable.
When users do not know what to say~\cite{HowDoUsersKnow}, our
facility to enquire about legal utterances may induce information overload
in large sites.  While we have not encountered this problem in
our PVS study, we suspect that applying out-of-turn interaction
in large web directories~(e.g.,~ODP) will involve new research directions.

\subsection{Acknowledgments}
We acknowledge the support
of NSF SGER Grant IIS-0136182 for funding this research.
We also extend our appreciation to the Virginia Tech
students who participated in our experiments.
We also thank Atul Shenoy~(Microsoft, Inc.) for assisting us with SALT
and Chris Williams (Virginia Tech CS) for
making the Extempore toolbar web installable for the study and
helping with screen captures.
Finally, we express our thanks to Srinidhi Varadarajan (Virginia Tech CS)
for recommending that we name our interface {\it Extempore}.

\bibliography{chi2004}

\begin{thebibliography}{10}

\bibitem{SALT}
{Speech Application Language Tags (SALT) Specification}.
\newblock Technical report, SALT Forum, July 2002.
\newblock Version 1.0.

\bibitem{MII-UR}
J.~F. Allen, C.~I. Guinn, and E.~Horvitz.
\newblock {Mixed-Initiative Interaction}.
\newblock {\em IEEE Intelligent Systems}, Vol. 14(5):pages 14--23,
  September--October 1999.

\bibitem{interactionScripts}
N.~J. Belkin, C.~Cool, A~Stein, and U.~Thiel.
\newblock {Cases, Scripts, and Information Seeking Strategies: On the Design of
  Interactive Information Retrieval Systems}.
\newblock {\em Expert Systems with Applications}, Vol. 9(3):pages 379--395,
  1995.

\bibitem{StrategyHubs}
S.~K. Bhavnani, C.~K. Bichakjian, T.~M. Johnson, R.~J. Little, F.~A. Peck,
  J.~L. Schwartz, and V.~J. Strecher.
\newblock {Strategy Hubs: Next-Generation Domain Portals with Search
  Procedures}.
\newblock In {\em Proceedings of the ACM Conference on Human Factors in
  Computing Systems (CHI'03)}, pages 393--400, Fort Lauderdale, FL, April 2003.
  ACM Press.

\bibitem{adaptiveHypermediaSurveyJoP2}
P.~Brusilovsky.
\newblock {Adaptive Hypermedia}.
\newblock {\em User Modeling and User-Adapted Interaction}, Vol. 11(1--2):pages
  87--110, 2001.

\bibitem{optimized-search}
S.~Dumais, E.~Cutrell, and H.~Chen.
\newblock {Optimizing Search by Showing Results in Context}.
\newblock In {\em Proceedings of the ACM Conference on Human Factors in
  Computing Systems (CHI'01)}, pages 277--284, Seattle, WA, April 2001. ACM
  Press.

\bibitem{Flamenco}
M.~A. Hearst, A.~Elliott, J.~English, R.~Sinha, K.~Swearingen, and K.-P. Yee.
\newblock {Finding the Flow in Web Site Search}.
\newblock {\em Communications of the ACM}, Vol. 45(9):pages 42--49, September
  2002.

\bibitem{ISinEE}
G.~Marchionini.
\newblock {\em {Information Seeking in Electronic Environments}}.
\newblock Cambridge Series on Human-Computer Interaction. Cambridge University
  Press, 1997.

\bibitem{info-scent}
C.~Olston and E.~H. Chi.
\newblock {ScentTrails: Integrating Browsing and Searching on the Web}.
\newblock {\em ACM Transactions on Computer-Human Interaction}, Vol.
  10(3):pages 177--197, September 2003.

\bibitem{MII-ITPro}
S.~Perugini and N.~Ramakrishnan.
\newblock {Personalizing Web Sites with Mixed-Initiative Interaction}.
\newblock {\em IEEE IT Professional}, Vol. 5(2):pages 9--15, March--April 2003.

\bibitem{pitkow-cacm}
J.~Pitkow, H.~Sch{\"{u}}tze, T.~Cass, R.~Cooley, D.~Turnbull, A.~Edmonds,
  E.~Adar, and T.~Breuel.
\newblock {Personalized Search}.
\newblock {\em Communications of the ACM}, Vol. 45(9):pages 50--55, September
  2002.

\bibitem{webAppSlicing}
F.~Ricca and P.~Tonella.
\newblock {Web Application Slicing}.
\newblock In {\em Proceedings of the International Conference on Software
  Maintenance (ICSM'01)}, pages 148--157, Florence, Italy, November 2001. IEEE
  Computer Society.

\bibitem{polyarchiesCHI}
G.~G. Robertson, K.~Cameron, M.~Czerwinski, and D.~Robbins.
\newblock {Polyarchy Visualization: Visualizing Multiple Intersecting
  Hierarchies}.
\newblock In {\em Proceedings of the ACM Conference on Human Factors in
  Computing Systems (CHI'02)}, pages 423--430, Minneapolis, MN, April 2002. ACM
  Press.

\bibitem{DynamicTax}
G.~M. Sacco.
\newblock {Dynamic Taxonomies: A Model for Large Information Bases}.
\newblock {\em IEEE Transactions on Knowledge and Data Engineering}, Vol.
  12(3):pages 468--479, May--June 2000.

\bibitem{HowDoUsersKnow}
N.~Yankelovich.
\newblock {How Do Users Know What To Say?}
\newblock {\em ACM Interactions}, 3(6):pages 32--43, November--December 1996.

\end{thebibliography}
\bibliographystyle{plain}
\end{document}